\documentclass[tightenlines,superscriptaddress,10pt,twocolumn,aps,prd]{revtex4-1}

\usepackage{geometry}
\usepackage{graphicx,wrapfig}
\usepackage[export]{adjustbox}
\usepackage{epsfig}
\usepackage{float}
\usepackage{flafter}
\usepackage{subfigure}
\usepackage{sidecap}
\usepackage{notes2bib}
\usepackage{gensymb}
\usepackage{url}
\usepackage{amsmath}
\usepackage{latexsym}
\usepackage{amssymb}
\usepackage{xcolor}
\usepackage{color}
\usepackage{slashed}
\usepackage{units}
\usepackage{siunitx}
\usepackage{notoccite}
\usepackage[utf8]{inputenc}
\usepackage[english]{babel}
\usepackage{csquotes}
\usepackage{dcolumn}
\usepackage{bm}
\usepackage{color}
\usepackage{enumitem}
\usepackage{pifont}
\usepackage{textcomp}
\usepackage{tikz}
\usepackage{colortbl}
\usepackage[version=4]{mhchem}
\usepackage{xspace}
\usepackage{placeins}
\usepackage{textcomp}
\usepackage{tabularx}
\usepackage{hyperref}
\usepackage{amssymb}
\usepackage{hyperref}
\usepackage{lineno}

\usepackage{pgfgantt}
\usepackage[ampersand]{easylist}

\newcommand{\cevns}{CEvNS}

\begin{document}

\onecolumngrid

\title{First Probe of Sub-GeV Dark Matter Beyond the Cosmological Expectation with the COHERENT CsI Detector at the SNS}

\widowpenalty10000
\clubpenalty10000
\renewcommand\floatpagefraction{1}
\renewcommand\topfraction{1}
\renewcommand\bottomfraction{1}
\renewcommand\textfraction{0}

\setlength{\belowcaptionskip}{-10pt} % tighten up float spacing

\renewcommand{\thesection}{\arabic{section}}
\renewcommand{\thesubsection}{\thesection.\arabic{subsection}}
\renewcommand{\thesubsubsection}{\thesubsection.\arabic{subsubsection}}

\makeatletter
\renewcommand{\p@subsection}{}
\renewcommand{\p@subsubsection}{}
\makeatother

\newcommand{\Mephidesc}{\affiliation{National Research Nuclear University MEPhI (Moscow Engineering Physics Institute), Moscow, 115409, Russian Federation}}
\newcommand{\Dukedesc}{\affiliation{Department of Physics, Duke University, Durham, NC, 27708, USA}}
\newcommand{\TUNLdesc}{\affiliation{Triangle Universities Nuclear Laboratory, Durham, NC, 27708, USA}}
\newcommand{\UTKdesc}{\affiliation{Department of Physics and Astronomy, University of Tennessee, Knoxville, TN, 37996, USA}}
\newcommand{\ITEPnewadesc}{\affiliation{National Research Center  ``Kurchatov Institute'' , Moscow, 123182, Russian Federation }}
\newcommand{\ORNLdesc}{\affiliation{Oak Ridge National Laboratory, Oak Ridge, TN, 37831, USA}}
\newcommand{\USDdesc}{\affiliation{Department of Physics, University of South Dakota, Vermillion, SD, 57069, USA}}
\newcommand{\NCSUdesc}{\affiliation{Department of Physics, North Carolina State University, Raleigh, NC, 27695, USA}}
\newcommand{\Sandiadesc}{\affiliation{Sandia National Laboratories, Livermore, CA, 94550, USA}}
\newcommand{\UWdesc}{\affiliation{Center for Experimental Nuclear Physics and Astrophysics \& Department of Physics, University of Washington, Seattle, WA, 98195, USA}}
\newcommand{\LANLdesc}{\affiliation{Los Alamos National Laboratory, Los Alamos, NM, 87545, USA}}
\newcommand{\Laurentiandesc}{\affiliation{Department of Physics, Laurentian University, Sudbury, Ontario, P3E 2C6, Canada}}
\newcommand{\CMUdesc}{\affiliation{Department of Physics, Carnegie Mellon University, Pittsburgh, PA, 15213, USA}}
\newcommand{\IUdesc}{\affiliation{Department of Physics, Indiana University, Bloomington, IN, 47405, USA}}
\newcommand{\VTdesc}{\affiliation{Center for Neutrino Physics, Virginia Tech, Blacksburg, VA, 24061, USA}}
\newcommand{\NCCUdesc}{\affiliation{Department of Mathematics and Physics, North Carolina Central University, Durham, NC, 27707, USA}}
\newcommand{\Tuftsdesc}{\affiliation{Department of Physics and Astronomy, Tufts University, Medford, MA, 02155, USA}}
\newcommand{\SNUdesc}{\affiliation{Department of Physics and Astronomy, Seoul National University, Seoul, 08826, Korea}}
\newcommand{\LLNLdesc}{\affiliation{Lawrence Livermore National Laboratory, Livermore, CA, 94550, USA}}
\newcommand{\WJCdesc}{\affiliation{Washington & Jefferson College, Washington, PA, 15301, USA}}
\author{D. Akimov}\Mephidesc
\author{P. An}\Dukedesc\TUNLdesc
\author{C. Awe}\Dukedesc\TUNLdesc
\author{P.S. Barbeau}\Dukedesc\TUNLdesc
\author{B. Becker}\UTKdesc
\author{V. Belov }\Mephidesc\ITEPnewadesc
\author{I. Bernardi}\UTKdesc
\author{M.A. Blackston}\ORNLdesc
\author{C. Bock}\USDdesc
\author{A. Bolozdynya}\Mephidesc
\author{J. Browning}\NCSUdesc
\author{B. Cabrera-Palmer}\Sandiadesc
\author{D. Chernyak}\altaffiliation{Now at:  Institute for Nuclear Research of NASU, Kyiv, 03028, Ukraine}\USDdesc\altaffiliation{Now at:  Institute for Nuclear Research of NASU, Kyiv, 03028, Ukraine}\altaffiliation{Now at:  Institute for Nuclear Research of NASU, Kyiv, 03028, Ukraine}
\author{E. Conley}\Dukedesc
\author{J. Daughhetee}\ORNLdesc
\author{J. Detwiler}\UWdesc
\author{K. Ding}\USDdesc
\author{M.R. Durand}\UWdesc
\author{Y. Efremenko}\UTKdesc\ORNLdesc
\author{S.R. Elliott}\LANLdesc
\author{L. Fabris}\ORNLdesc
\author{M. Febbraro}\ORNLdesc
\author{A. Gallo Rosso}\Laurentiandesc
\author{A. Galindo-Uribarri}\ORNLdesc\UTKdesc
\author{M.P. Green}\TUNLdesc\ORNLdesc\NCSUdesc
\author{M.R. Heath}\ORNLdesc
\author{S. Hedges}\Dukedesc\TUNLdesc\LLNLdesc
\author{D. Hoang}\CMUdesc
\author{M. Hughes}\IUdesc
\author{T. Johnson}\Dukedesc\TUNLdesc
\author{A. Khromov}\Mephidesc
\author{A. Konovalov}\Mephidesc\ITEPnewadesc
\author{E. Kozlova}\Mephidesc\ITEPnewadesc
\author{A. Kumpan}\Mephidesc
\author{L. Li}\Dukedesc\TUNLdesc
\author{J.M. Link}\VTdesc
\author{J. Liu}\USDdesc
\author{K. Mann}\NCSUdesc
\author{D.M. Markoff}\NCCUdesc\TUNLdesc
\author{J. Mastroberti}\IUdesc
\author{P.E. Mueller}\ORNLdesc
\author{J. Newby}\ORNLdesc
\author{D.S. Parno}\CMUdesc
\author{S.I. Penttila}\ORNLdesc
\author{D. Pershey}\Dukedesc
\author{R. Rapp}\altaffiliation{Now at: Washington \& Jefferson College, Washington, PA, 15301, USA}\CMUdesc\altaffiliation{Now at: Washington \& Jefferson College, Washington, PA, 15301, USA}\altaffiliation{Now at: Washington \& Jefferson College, Washington, PA, 15301, USA}
\author{J. Raybern}\Dukedesc
\author{O. Razuvaeva}\Mephidesc\ITEPnewadesc
\author{D. Reyna}\Sandiadesc
\author{G.C. Rich}\TUNLdesc
\author{J. Ross}\NCCUdesc\TUNLdesc
\author{D. Rudik}\Mephidesc
\author{J. Runge}\Dukedesc\TUNLdesc
\author{D.J. Salvat}\IUdesc
\author{A.M. Salyapongse}\CMUdesc
\author{J. Sander}\USDdesc
\author{K. Scholberg}\Dukedesc
\author{A. Shakirov}\Mephidesc
\author{G. Simakov}\Mephidesc\ITEPnewadesc
\author{G. Sinev}\altaffiliation{Now at: South Dakota School of Mines and Technology, Rapid City, SD, 57701, USA}\Dukedesc\altaffiliation{Now at: South Dakota School of Mines and Technology, Rapid City, SD, 57701, USA}
\author{W.M. Snow}\IUdesc
\author{V. Sosnovtsev}\Mephidesc
\author{B. Suh}\IUdesc
\author{R. Tayloe}\IUdesc
\author{K. Tellez-Giron-Flores}\VTdesc
\author{I. Tolstukhin}\altaffiliation{Now at: Argonne National Laboratory, Argonne, IL, 60439, USA}\IUdesc\altaffiliation{Now at: Argonne National Laboratory, Argonne, IL, 60439, USA}
\author{E. Ujah}\NCCUdesc\TUNLdesc
\author{J. Vanderwerp}\IUdesc
\author{R.L. Varner}\ORNLdesc
\author{C.J. Virtue}\Laurentiandesc
\author{G. Visser}\IUdesc
\author{T. Wongjirad}\Tuftsdesc
\author{Y.-R. Yen}\CMUdesc
\author{J. Yoo}\SNUdesc
\author{C.-H. Yu}\ORNLdesc
\author{J. Zettlemoyer}\altaffiliation{Now at: Fermi National Accelerator Laboratory, Batavia, IL, 60510, USA}\IUdesc\altaffiliation{Now at: Fermi National Accelerator Laboratory, Batavia, IL, 60510, USA}

%\linenumbers

\begin{abstract}

The COHERENT collaboration searched for scalar dark matter particles produced at the Spallation Neutron Source with masses between 1 and 220~MeV/c$^2$ using a CsI[Na] scintillation detector sensitive to nuclear recoils above 9~keV$_\text{nr}$.  No evidence for dark matter is found and we thus place limits on allowed parameter space.  With this low-threshold detector, we are sensitive to coherent elastic scattering between dark matter and nuclei.  The cross section for this process is orders of magnitude higher than for other processes historically used for accelerator-based direct-detection searches so that our small, 14.6~kg detector significantly improves on past constraints.  At peak sensitivity, we reject the flux consistent with the cosmologically observed dark-matter concentration for all coupling constants $\alpha_D<0.64$, assuming a scalar dark-matter particle.  We also calculate the sensitivity of future COHERENT detectors to dark-matter signals which will ambitiously test multiple dark-matter spin scenarios.

\end{abstract}

\maketitle

%\clearpage
%\newpage
%\mbox{~}
%\clearpage
%\newpage

\twocolumngrid
\textit{Introduction:} Standard model (SM) fermions only account for $\approx20\%$ of cosmologically observed matter \cite{Freese:2017idy}.  The remaining matter, called dark matter (DM), was postulated nearly 100 years ago to explain anomalous orbital speeds of stars within galaxies~\cite{1939LicOB..19...41B} and galaxies within clusters~\cite{Zwicky:1933gu,Zwicky:1937zza}.  Despite continuing improvement in our understanding of the gravitational effects of DM, its particle nature has not been determined.  Searches for traditional weakly interacting massive particle (WIMP) DM have not yet found a positive signature \cite{PhysRevLett.121.111302,PhysRevLett.123.251801,XENON:2020fgj,PhysRevLett.121.111303,PhysRevLett.127.061801}.  Further, experimental sensitivity is rapidly approaching a ``neutrino fog" \cite{Vergados:2008jp} of background from coherent elastic neutrino-nucleus scattering~(\cevns{}) \cite{PhysRevD.9.1389} events from astrophysical neutrino sources which will hinder progress.  

In response, the interest in sub-GeV DM particles, too light to be observed in many conventional WIMP detectors, has increased recently.  Cosmological observations suggest that such DM could not interact with SM matter through the weak force~\cite{PhysRevLett.39.165}.  However, sub-GeV hidden-sector DM particles could interact with SM fermions mediated by a ``portal" particle \cite{PhysRevD.70.023514,BOEHM2004219,Pospelov:2007mp,PhysRevD.84.075020}.  These proposed hidden sector particles are viable DM candidates.  

If sub-GeV DM exists, these particles would be produced at accelerators.  Beam-dump experiments have already begun to survey the possible parameter space \cite{PhysRevD.63.112001,Aguilar-Arevalo:2017mqx,PhysRevD.98.112004,CCM:2021leg,BaBar:2008aby,NA64:2019imj,Batell:2014mga} with more experiments planned \cite{Akesson:2018vlm,Battaglieri:2020lds}.  Searches for accelerator-produced DM are of particular interest as the DM particles are relativistic so that the scattering cross section is relatively spin-independent \cite{Battaglieri:2017aum}.

Experiments capable of seeing low-energy nuclear recoils associated with \cevns{} can search for an analogous coherent elastic DM-nucleus scattering process \cite{deNiverville:2015mwa,Dutta:2020vop}.  As the cross section scales according to the square of the proton number $Z^2$, such an experiment can achieve competitive sensitivity with relatively low mass.  Further, as sub-GeV interactions with the SM are mediated by a new force particle, the hadronic and leptonic DM couplings may be radically different, thus calling for experimental efforts to constrain both.  While the majority of previous constraints test the coupling between DM and leptons, CEvNS experiments are sensitivity to coherent DM-nucleus scattering, thus probing the quark coupling.

In this paper, we present the first search by COHERENT for accelerator-produced DM particles.  This uses data collected by our CsI detector which measured \cevns{} \cite{Akimov:2017ade,COHERENT:2021xmm} at the Spallation Neutron Source (SNS) at Oak Ridge National Laboratory~\cite{Mason:2000wb}.  We focus on a single benchmark model of scalar DM particle, $\chi$, mediated by a vector portal particle, $V$, \cite{PhysRevD.84.075020} with masses $m_\chi$ and $m_V$, respectively.  In this model, $V$ kinetically mixes with the SM photon with a coupling $\varepsilon$.  DM particles can then be produced through $V\rightarrow\chi\bar{\chi}$ decay with a coupling $\alpha_D$.  The DM scatterign cross section at thermal freeze-out, and thus relic DM abundance in the modern universe, depends on the single parameter, $Y$~\cite{Izaguirre:2015yja}, defined as
\begin{equation}
    Y = \varepsilon^2\alpha_D\left(\frac{m_\chi}{m_V}\right)^4.
\end{equation}
We thus adopt this parameter when presenting our results for convenient comparison to other measurements.  Our analysis is restricted to scalar DM in the mass range $1$~$<$~$m_\chi$~$<$~$220$~MeV$/c^2$.  A spin~\textonehalf~particle is also viable as a DM candidate, though these scenarios would require lower couplings to match the cosmologically observed concentration.  With improved sensitivity in the future, we will explore constraints on Majorana and pseudo-Dirac fermion DM, but currently focus on scalar DM.

\textit{The COHERENT CsI detector at the SNS:} The SNS operates a 1.4~MW proton beam incident on a mercury target running at 60~Hz.  For our detector operations, the SNS maintained a beam-pulse width of 378~ns FWHM and an average proton energy of 0.984~GeV.  With its high beam power, the SNS could produce an enormous flux of DM particles through proton bremsstrahlung and hidden-sector decays of $\pi^0$ and $\eta^0$ mesons produced in the target.  Neutrinos from the accelerator, produced by the decay of $\pi^+$ particles which formed as protons stop in the target, induce \cevns{} in our detectors, one of our principal backgrounds to dark-matter detection.  This neutrino flux includes a prompt component from $\pi^+\rightarrow\mu^+\nu_\mu$ and a delayed component, $\tau=2.2\mu$s, from $\mu^+\rightarrow e^+\nu_e\bar{\nu}_\mu$ decay.

We operate several detectors in ``Neutrino Alley" at the SNS, a basement hallway with sufficiently low backgrounds to allow for neutrino measurements.  One of our detectors was a 14.6~kg CsI[Na] scintillating crystal \cite{Collar:2014lya,2014PhDT.......121F}, commissioned in 2015, which made the first observation of \cevns{} \cite{Akimov:2017ade}.  This detector was decommissioned in 2019.  During its run, this detector collected 13.99~GWhr, $3.20\times10^{23}$~protons on target, of beam data.  The detector was situated 19.3~m from the beam target, 90$^\circ$ off-axis from the beam direction.  The light was collected by a single Hamamatsu R877-100 photomultiplier (PMT) sampled at a rate of 500~MS/s.  We assembled shielding with multiple materials to moderate both environmental $\gamma$ and neutron activity.

The light yield of the detector was determined to be 13.35 photo-electrons (PE) / keV-electron-equivalent (keV$_\text{ee}$).  Two sources were used for calibration: a 59~keV$_\text{ee}$ $\gamma$ line from $^{60}$Co decay and a 57.5~keV$_\text{ee}$ peak from $^{127}$I($n$,$\gamma$) which includes a small, quenched nuclear recoil.  The light yield was shown to be uniform across the crystal by taking calibration data at multiple locations along the detector length.

\textit{Dark matter events in the CsI detector:} We use the BdNMC \cite{deNiverville:2016rqh} simulation package to predict the DM flux in Neutrino Alley along with the scattering rate and kinematics within our detectors.  BdNMC is versatile, calculating DM production and detection through several channels.  Coherent elastic DM-nucleus scattering has been implemented specifically for \cevns{} experiments.  

The dominant production channels for portal particles at the SNS are $\pi^0\rightarrow\gamma+V$ decay, $\eta^0\rightarrow\gamma+V$ decay, and $p+N\rightarrow p+N+V$ bremsstrahlung.  Production from $\pi^0$ decay, $\eta^0$ decay, and proton bremsstrahlung dominate for DM masses below 40~MeV/c$^2$, between 40 and 130~MeV/c$^2$, and above 130~MeV/c$^2$, respectively.  We do not have sensitivity for $m_\chi>220$~MeV/c$^2$, beyond which bremsstrahlung is kinematically forbidden.  With a GEANT4 \cite{Geant4Cite} simulation, we predict $0.107$~$\pm10\%$~$\pi^0$ produced per proton incident on the target~\cite{PhysRevD.106.032003}.  For our most sensitive masses, $\pi^0$ decay dominates the sensitivity.  The production of portal particles from $\pi^0$ decay is given by the branching ratio
\begin{equation}
    \frac{\text{Br}(\pi^0\rightarrow\gamma V)}{\text{Br}(\pi^0\rightarrow\gamma\gamma)}=2\varepsilon^2\left(1-\frac{m_V^2}{m_\pi^2}\right)^3
\end{equation}
for $m_V<m_\pi$~\cite{Batell:2014yra} which is proportional to the expected DM flux.  Though the beam energy at the SNS, $T_p\approx0.98$~GeV, is slightly lower than the production threshold for $p+p$~$\rightarrow$~$p+p+\eta^0$ production, there are $\eta$ mesons produced in the target due to the Fermi momentum of mercury~\cite{subthresholdMeasurement}.  A calculation of this sub-threshold production \cite{Cassing1991} suggests that about $0.002$~$\pm30\%$~$\eta^0$ are produced per $\pi^0$ at the SNS.  BdNMC predicts the timing of scattering events which typically scatter within a few~ns of the speed-of-light-delayed DM production in the target.  As this is a small delay, we assume all DM we study travels at the speed of light.

\begin{figure}[!bt]
\centering
\includegraphics[width=0.49\textwidth]{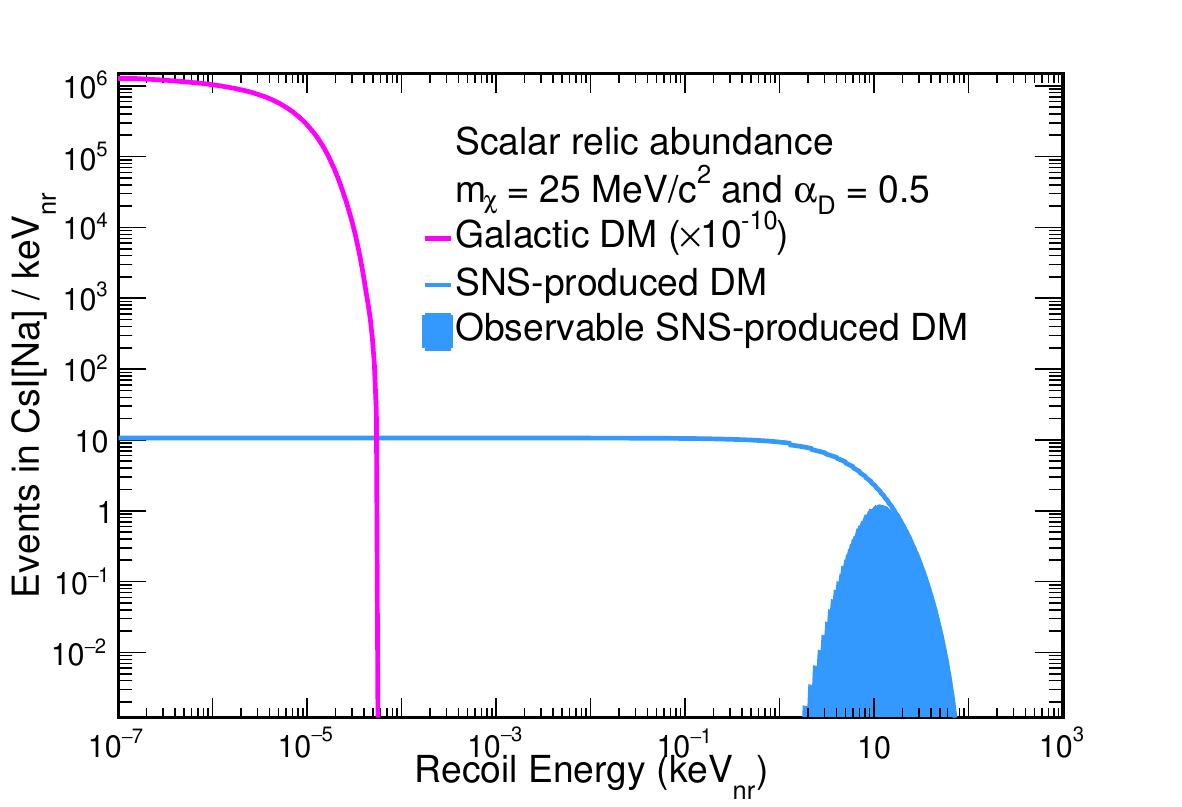}
\caption{The expected spectrum of coherent DM scattering signal in CsI from both galactic and SNS-produced DM for a mass of 25~MeV/c$^2$, near our optimal sensitivity.  Though the rate for galactic DM is higher, the recoil energies are far below threshold while we select 26$\%$ of DM produced at the SNS.}
\label{fig:DMEff}
\end{figure}

To lowest order, the differential cross section in recoil energy, $E_r$, is
\begin{equation}
    \frac{d\sigma}{dE_r} = 4\pi\alpha_D\alpha\varepsilon^2Z^2\frac{2m_NE_\chi^2}{p_\chi^2(m_V^2+2m_NE_r)^2}
    \label{eqn:DMCohXSec}
\end{equation}
where $\alpha$ is the electromagnetic fine structure constant, $m_N$ is the nuclear mass, and $p_\chi$ and $E_\chi$ are the incident DM momentum and energy.  The event rate is given by the flux $\times$ the cross section and thus depends on the couplings as $\propto\varepsilon^4\alpha_D\propto Y^2/\alpha_D$.  The scattering model used for our sensitivity estimates presented in \cite{Akimov:2019xdj} had a calculation error with the definition of $Q^2=2m_NE_r$, described in~\cite{deNiverville:2015mwa}, that has now been fixed.  We have confirmed event rates predicted by BdNMC using this new model with an independent, cross-check calculation from COHERENT.  

For $m_\chi=25$~MeV/$c^2$, the expected average recoil energy is 9~keV, just at our analysis threshold.  The spectra of interacting DM in our CsI detector are shown in Fig.~\ref{fig:DMEff} for both galactic and SNS-produced DM assuming a DM mass at our peak sensitivity, $m_\chi$~$=$~$25$~MeV$/c^2$.  The prediction of the galactic recoil spectrum assumes a local DM density of 0.3~GeV/cm$^3$ near Earth with a Boltzmann speed distribution with $v/c\approx0.001$~\cite{Lacroix_2020}.  Though fewer interactions are expected for SNS-produced DM, the typical recoil energy is higher than for galactic DM by a factor of 10$^6$ allowing for detection of 26$\%$ of the SNS-produced DM.

The detector response for DM recoil events is assumed to be the same as for \cevns{}, described in \cite{COHERENT:2021xmm}, apart from quenching at high recoil energies.  All data used to fit our quenching model were taken at $E_\text{rec}<70$~keV$_\text{nr}$.  This is sufficient to cover all \cevns{} recoils; however a small percentage of DM-induced recoils lie beyond this point.  For recoil energies above 70~keV$_\text{nr}$, we assume a constant quenching factor, $(9.8\pm1.8)\%$, which is the quenching and uncertainty implied by our fit at 70~keV$_\text{nr}$.

\textit{Data analysis:} We performed a search for light DM particles in our CsI data collected during SNS operations.  The analysis was blinded, defining all selection cuts, uncertainties, and fitting methods before determining the observed data spectrum.  The DM scattering model, however, was updated after box-opening to correct an error discovered in the coherent cross section.  The corrected version is given by Eqn.~\ref{eqn:DMCohXSec}.  

We used the same event reconstruction used to determine the \cevns{} cross section \cite{COHERENT:2021xmm}.  We also applied the same event selection, except that the highest recoil energy analyzed was increased from 60~PE to 250~PE to capture the most energetic recoils expected from high-mass DM interactions.  The recoil energy binning was also re-optimized for ideal DM separation from the backgrounds.  The analysis binning was not reoptimized after box opening and updating the cross section model.  Steady-state accidentals (SSBkg) and \cevns{} interactions are the dominant backgrounds.  A small number of beam-related neutron (BRN) and neutrino-induced neutron (NIN) events were also accounted for.  Neutron rates and uncertainties were determined from EJ-301 \footnote{Eljen Technology, 1300 W. Broadway St., Sweetwater, TX 79556} liquid scintillator data collected before detector commissioning \cite{COHERENT:2021xmm}.

The \cevns{} cross section was fixed to the SM prediction and allowed to float within the form-factor uncertainty, 3.4$\%$.  We also included systematic uncertainties from neutrino flux, background normalization, threshold, selection efficiency, and quenching that are calculated as described in \cite{COHERENT:2021xmm,Akimov_2022} and propagated to the DM signal prediction when appropriate.  The form-factor, efficiency, and quenching uncertainties included both normalization and shape uncertainties.

Our DM prediction, parameterized by DM mass, $m_\chi$, and coupling, $Y$, was added to our expected SSBkg, BRN, NIN, and \cevns{} backgrounds.  We tested DM masses between 1 and 220~MeV$/c^2$ which covers the range where COHERENT has competitive sensitivity.

The timing of observed events, $t_\text{rec}$, is critical for this result.  We measure $t_\text{rec}$ relative to a trigger signal from the accelerator indicating protons on target, and the neutrino pulse arrives in our detector roughly 150~ns later.  The DM region of interest (ROI) is defined as $0.25 \leq t_\text{rec} < 0.75$~$\mu$s, determined by selecting the optimal timing region for separating the DM signal from backgrounds with a $s/\sqrt{b}$ figure of merit.  Over 92$\%$ of DM recoils but only 25$\%$ of \cevns{} events are expected in this interval.  Most neutrino events are delayed relative to the DM events by $\tau_\mu=2.2$~$\mu$s.  These delayed events can be used to constrain systematic uncertainties in situ to improve the precision of background estimates within the DM ROI.

%We assume $\alpha_D=0.5$ with lower values of $\alpha_D$ giving more strict constraints while values $\alpha_D>1$ yield a nonperturbative model.  We also assume $m_V/m_\chi=3$ throughout, again a conservative choice~\cite{deNiverville:2015mwa}.  

The data was binned in two dimensions: recoil energy and recoil time.  For each DM mass and coupling, we performed a binned log-likelihood fit profiling over nuisance parameters relating to systematic uncertainties.  As $Y$ depends on two couplings, $\alpha_D$ and $\varepsilon$, we conservatively fix $\alpha_D=0.5$ in our constraint~\cite{deNiverville:2015mwa}.  Lower values reduce both the expected event rate and $Y$.  Since the DM scattering cross section at freeze-out is $\propto Y$, when assuming a fixed DM relic abundance, a decrease in $\alpha_D$ must correspond to an increase in $\varepsilon$ such that the total event rate increases.  Thus, lower values of $\alpha_D$ give tighter bounds on the dark matter model.  For $\alpha_D > 0.5$, the scattering cross section, and thus event rate, increases due to higher order diagrams, thus improving constraints.  $Y$ also depends on two masses, $m_\chi$ and $m_V$.  As the production and scattering both only depend on $m_V$, increasing the value of $m_V/m_\chi$ with a fixed $m_V$ does not affect our event rate but does give lower values of $Y$ and tighter bounds.  We thus assume $m_V/m_\chi=3$ according to convention near the threshold for on-shell $V$ decay though there is viable parameter space at lower values.  For a given value of $m_\chi$, we calculate the $\Delta\chi^2(Y)$ curve relative to the best-fit DM coupling.  Allowed values of $Y$ are determined according to the Feldman-Cousins prescription \cite{Feldman:1997qc} with 90$\%$ confidence.

\begin{figure}[!bt]
\centering
\includegraphics[width=0.49\textwidth]{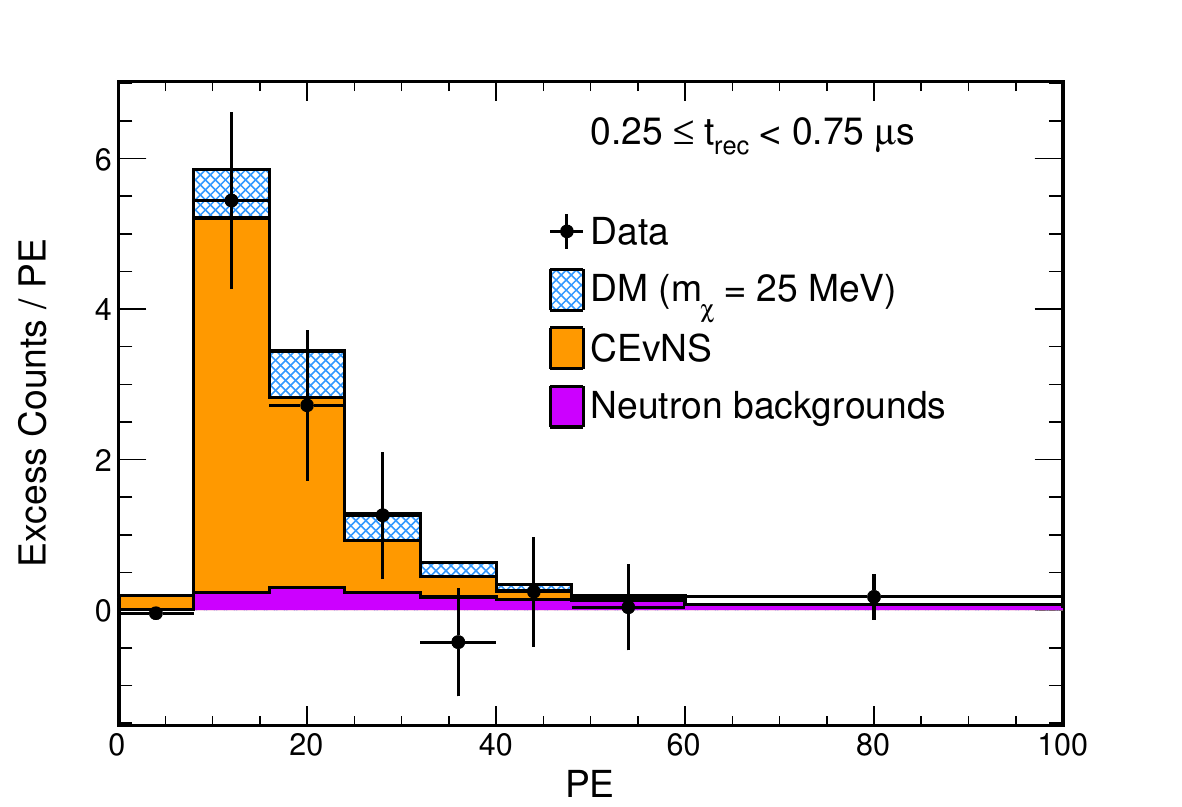}
\includegraphics[width=0.49\textwidth]{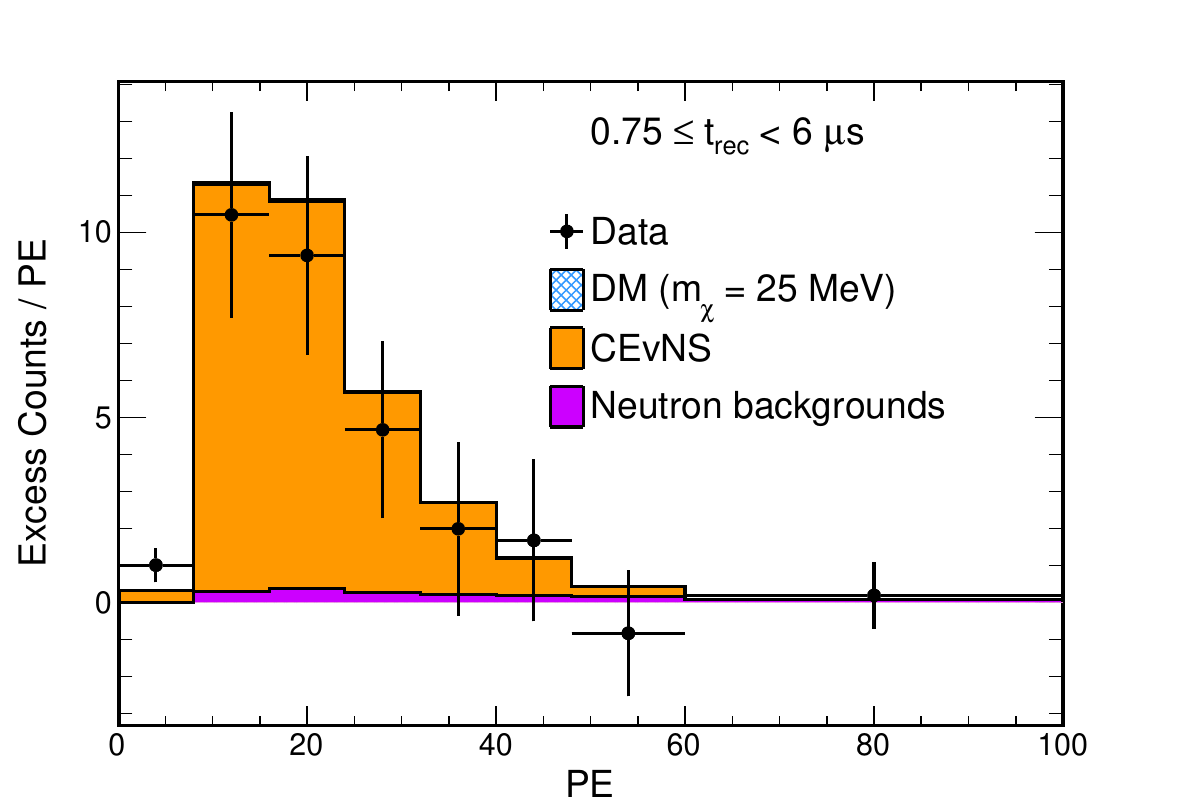}
\caption{The observed, SSBkg-subtracted recoil spectra in the DM timing ROI (top) and the background control sample (bottom) compared to the best-fit prediction with no DM.  The expected DM distribution at the 90$\%$ limit is stacked on the SM prediction for $m_\chi$~$=$~$25$~MeV/c$^2$.}
\label{fig:DMSpectra}
\end{figure}

\textit{Results:} We selected 5142 events in the analysis region of $0\leq E_\text{rec}<250$~PE and $0\leq t_\text{rec}<6$~$\mu$s.  For each DM mass tested, the best-fit was identical, preferring no DM events in each case with a fit $\chi^2/\text{dof}=103.0/120$.  Our observed best-fit, SSBkg-subtracted spectra, both in the DM timing ROI and the \cevns{} background timing region, are shown in Fig.~\ref{fig:DMSpectra} with our 90$\%$ limit on DM events.  A summary of background counts in the sample is shown in Tab.~\ref{table:EventTotals}.

\begin{figure}[!bt]
\centering
\includegraphics[width=0.49\textwidth]{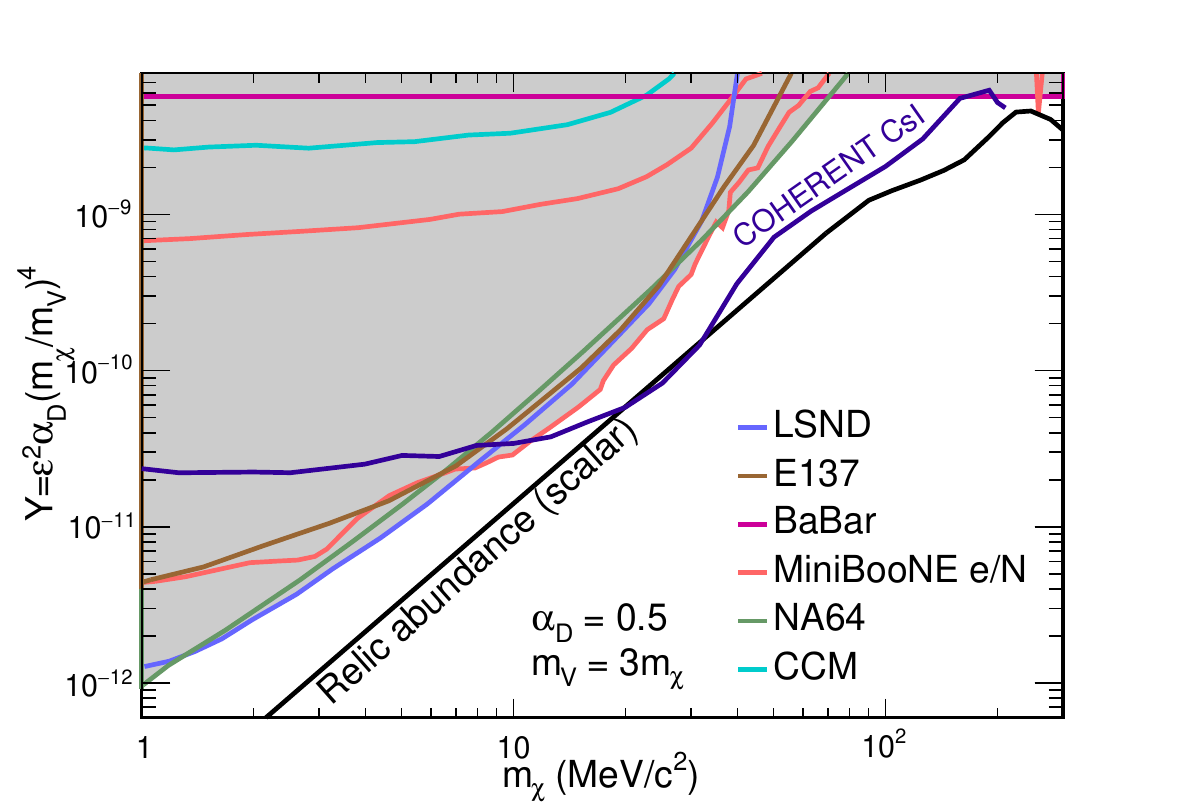}
\includegraphics[width=0.49\textwidth]{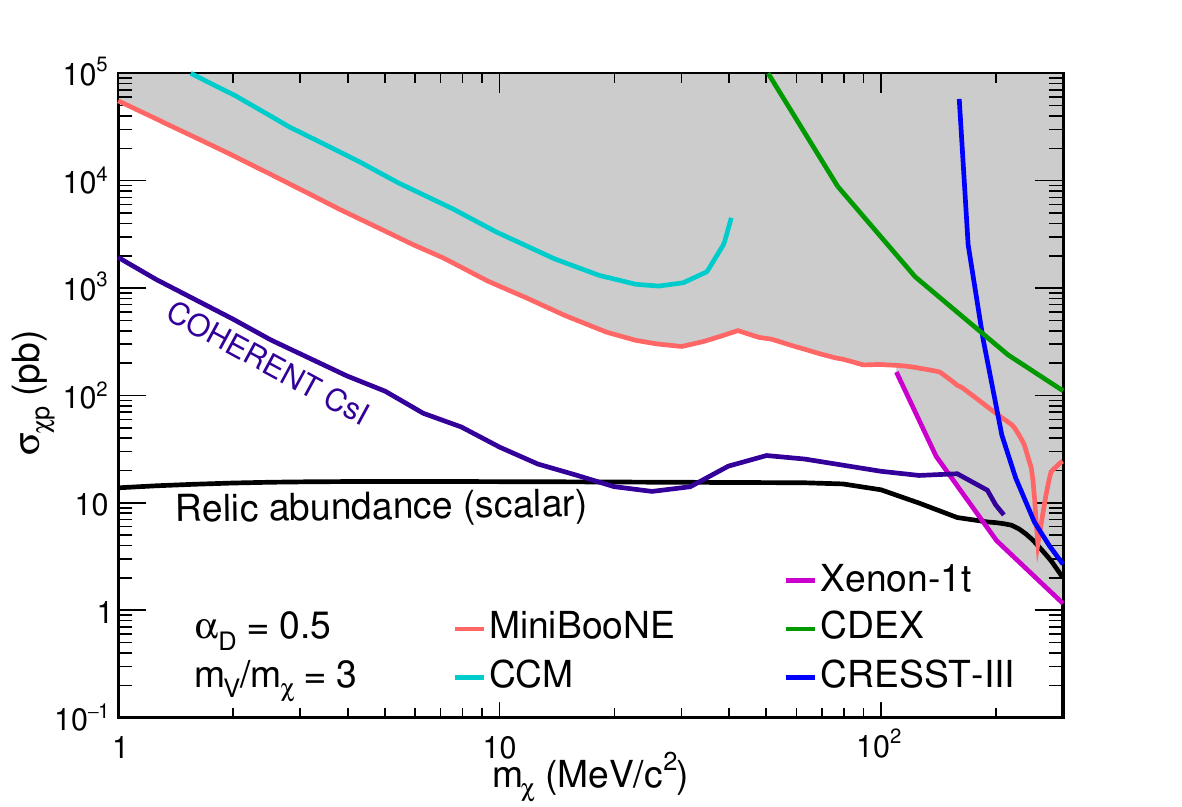}
\caption{Constraint of DM parameter space for COHERENT CsI data compared to other experimental data, assuming $\alpha_D=0.5$ (top).  The parameters which give the observed relic abundance for scalar DM are also shown.  We also show constraints on the DM-nucleon scattering cross section, averaged over the DM halo velocity distribution compared to constraints of the DM-quark coupling of light DM (bottom).}
\label{fig:Result}
\end{figure}

\begin{figure}[!bt]
\centering
\includegraphics[width=0.49\textwidth]{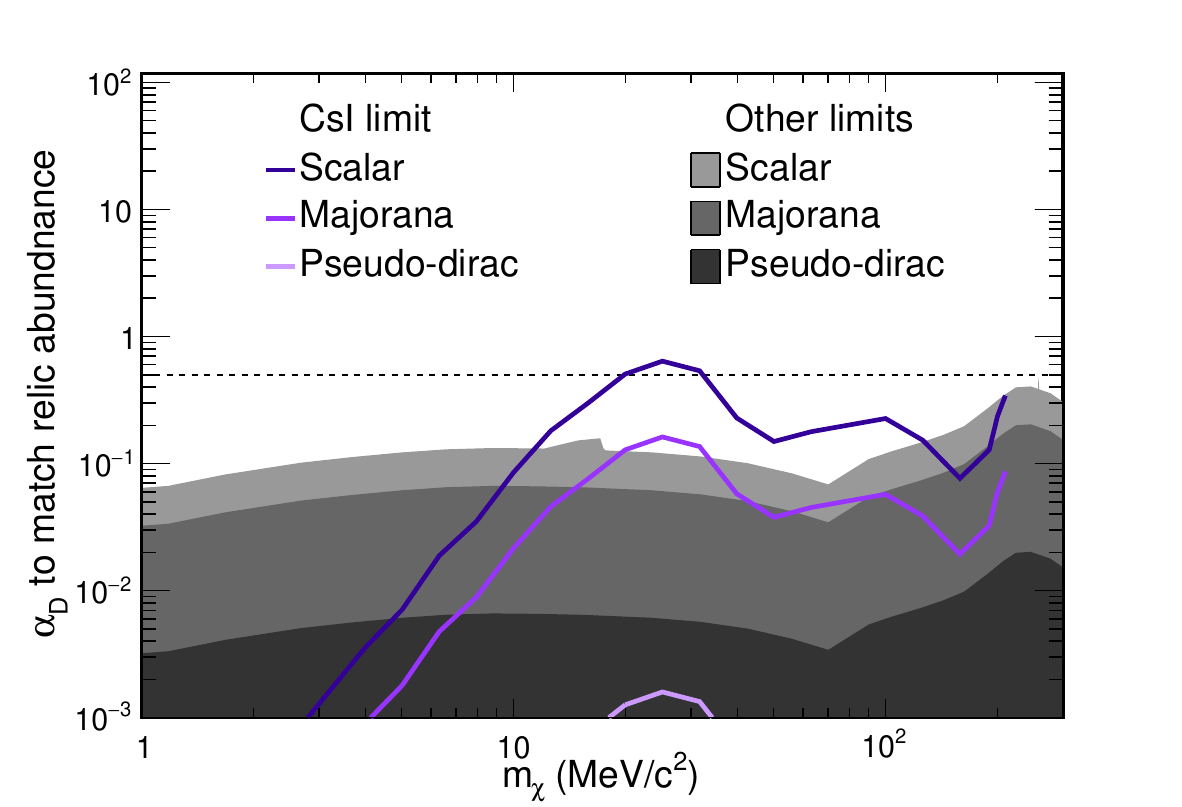}
\caption{The values of $\alpha_D$ for which we can reject the cosmologically observed DM concentration as a function of DM mass compared to constraints shown in Fig.~\ref{fig:Result}.  All parameter space below the contour is excluded.  We include scenarios where DM is scalar, a Majorana fermion, or a pseudo-Dirac fermion.  A dashed line at $\alpha_D=0.5$, the value assumed in Fig.~\ref{fig:Result}, is also drawn.}
\label{fig:AlphaResult}
\end{figure}

\begin{table}[H]
\begin{center}
\begin{tabular}{c|cc}
 & Prior Prediction & Best-Fit Total \\
\hline
SSBkg & $4893\pm70$ & $4857\pm62$ \\
BRN & $27.6\pm6.9$ & $25.8\pm6.7$ \\
NIN & $7.6\pm2.7$ & $7.4\pm2.7$ \\
\cevns{} & $341\pm36$ & $320\pm32$ \\
DM & $-$ & $<15.8$ \\
\end{tabular}
\caption{A summary of prior prediction and best-fit event rates for each background and the 90$\%$ limit for 25~MeV/c$^2$ DM.}
\label{table:EventTotals}
\end{center}
\end{table}

The fit prefers slightly fewer \cevns{} events than predicted after profiling over nuisance parameters.  This is consistent with our \cevns{} measurement using the same dataset \cite{COHERENT:2021xmm}.  Our critical $\Delta\chi^2$ values used to construct 90$\%$ confidence intervals on $N_\text{DM}$ are $2.1-2.3$ depending on $m_\chi$.  These are slightly lower than those expected from Gaussian statistics due to the proximity to the boundary $N_\text{DM}\geq0$.  At our peak sensitivity, $m_\chi=25$~MeV/c$^2$, we determined there are $<15.8$ DM events in our sample to $90\%$ CL, though the constraint on the number of DM scatters in our dataset depends on the mass assumption.

Our constraint on DM parameters for $\alpha_D=0.5$ is shown in Fig.~\ref{fig:Result} along with our projected sensitivity and other current constraints.  The parameters that yield the relic abundance for scalar DM \cite{Dutta:2019eml} is also shown.  With a small 14.6~kg detector, we improve constraints on $Y$ for $11$~$<$~$m_\chi$~$<$~$165$~MeV/c$^2$ by up to $5\times$ suggesting that future, large-scale \cevns{} detectors will be successful in ambitiously limiting light DM models.  With the current dataset, we can reject coupling parameters consistent with cosmological DM for masses between 20 and 33~MeV/c$^2$ assuming $\alpha_{D}=0.5$.  The constraint is strongest at $m_\chi=25$~MeV/c$^2$ where we can eliminate the scalar target for all $\alpha_D<0.64$.

Additionally, as there are few accelerator-based searches for DM that test the DM-quark coupling, we also compare our constraint to both astroparticle and accelerator-based searches of light DM sensitive to the quark coupling.  Comparisons to astroparticle results are made by averaging the coherent DM-nucleus cross section given in Eqn.~\ref{eqn:DMCohXSec} with couplings determined by our constraint over the velocity distribution expected for the DM halo near Earth~\cite{Lacroix_2020}.  This result is also shown in Fig.~\ref{fig:Result}.  Our constraint improves on all constraints of the DM-quark coupling for masses below 166~MeV/$c^2$ where COHERENT data probes more than an order of magnitude of previously untested parameter space.  At higher masses, astroparticle experiments exploiting the Migdal effect~\cite{migdalOriginal,Vergados:2005dpd,Ibe:2017yqa} dominate~\cite{PhysRevLett.123.241803,PhysRevLett.123.161301} with an additional constraint from CRESST-III~\cite{CRESST:2019jnq}.

As our constraint depends on our particular choice of $\alpha_D$, we can explore this parameter by constraining the values of $\alpha_D$ for which we reject the relic abundance at 90$\%$, as shown in Fig.~\ref{fig:AlphaResult}.  For a given DM mass, the relic abundance is given by a fixed value of $Y$.  Decreasing $\alpha_D$ while holding $Y$ fixed at the relic abundance increases $\varepsilon\propto1/\sqrt{\alpha_D}$ such that the overall signal rate expected in COHERENT, which scales like $\varepsilon^4\alpha_D\propto1/\alpha_D$, increases.  Thus, we show the lower bounds for allowed $\alpha_D$.  For scalar DM, we constrain the cosmological abundance with very conservative choices of $\alpha_D$.  However, if DM is a Majorana or a pseudo-Dirac fermion, significant parameter space remains.  In the future, with larger detectors sensitive to lower nuclear recoils, \cevns{} data can probe fermion DM models at $\alpha_D\approx0.5$, which favor $Y$ values up to $20\times$ lower.

\textit{Conclusion:} We have exploited \cevns{} data collected by our decommissioned CsI[Na] detector at the SNS to search for hidden-sector DM particles produced in the beam.  This dataset, in addition to making the most precise measurement of \cevns{} to date, has considerably improved on current constraints for accelerator-produced DM particles with masses between 11 and 165~MeV/c$^2$.  Additionally, this result improves on constraints of the DM-quark coupling for masses below 166 MeV/$c^2$.  In particular, this is the first result to test scalar DM for even the conservative choice of $\alpha_D=0.5$ in the studied mass range.  The data also constrains Majorana and pseudo-Dirac DM, but constraints on these scenarios will not be as exhaustive until future data is collected.  We have developed powerful methods for understanding background rates by exploiting timing information.  In the future, these techniques will constrain systematic uncertainties in-situ allowing much more stringent searches.  In particular, future argon and CsI detectors placed in the STS beam have significant potential to discover an excess of DM scatters in currently unexplored parameter space independent of DM mass and spin phenomenology.

\textit{Acknowledgements:} The COHERENT collaboration acknowledges the Kavli Institute at the University of Chicago for CsI[Na] detector contributions.  We are very grateful for helpful conversations with P. deNiverville, B. Dutta, D. Kim, L. Strigari, and A. Thompson for determining our scattering model.  The COHERENT collaboration acknowledges the resources generously provided by the Spallation Neutron Source, a DOE Office of Science User Facility operated by the Oak Ridge National Laboratory. This work was supported by the US Department of Energy (DOE), Office of Science, Office of High Energy Physics and Office of Nuclear Physics; the National Science Foundation; the Consortium for Nonproliferation Enabling Capabilities; the Consortium for Monitoring Technology and Verification, the Ministry of Science and Higher Education of the Russian Federation (Project ``Fundamental properties of elementary particles and cosmology'' No. 0723-2020-0041); the Russian Foundation for Basic Research (Project 20-02-00670\_a); and the US DOE Office of Science Graduate Student Research (SCGSR) program, administered for DOE by the Oak Ridge Institute for Science and Education which is in turn managed by Oak Ridge Associated Universities. Sandia National Laboratories is a multi-mission laboratory managed and operated by National Technology and Engineering Solutions of Sandia LLC, a wholly owned subsidiary of Honeywell International Inc., for the U.S. Department of Energy's National Nuclear Security Administration under contract DE-NA0003525. The Triangle Universities Nuclear Laboratory is supported by the U.S. Department of Energy under grant DE-FG02-97ER41033. Laboratory Directed Research and Development funds from Oak Ridge National Laboratory also supported this project. This research used the Oak Ridge Leadership Computing Facility, which is a DOE Office of Science User Facility.

%\bibliography{main.bib}
\bibliography{main.bbl}

\end{document}